\documentclass{article}

\usepackage{arxiv}

\usepackage[utf8]{inputenc} 
\usepackage[T1]{fontenc}    
\usepackage{hyperref}       
\usepackage{url}            
\usepackage{booktabs}       
\usepackage{amsfonts}       
\usepackage{nicefrac}       
\usepackage{microtype}      
\usepackage{lipsum}		
\usepackage{graphicx}
\usepackage{doi}
\usepackage{csquotes}
\usepackage{adjustbox}
\usepackage{subcaption}
\usepackage{array}
\usepackage{tabularx}
\usepackage{titlesec}
\usepackage{multirow}
\usepackage{wrapfig}
\newcommand\st[1]{\footnotesize\itshape{}{#1}}

\title{\enquote{What can I cook with these ingredients?} - Understanding cooking-related information needs in conversational search}

\author{ 
    {Alexander Frummet}\\
	Universität Regensburg\\
	Universitätsstraße 31\\
	Regensburg, Germany \\
	\texttt{alexander.frummet@ur.de} 
	\And
	{David Elsweiler}\\
	Universität Regensburg\\
	Universitätsstraße 31\\
	Regensburg, Germany \\
	\texttt{david.elsweiler@ur.de} 
	\And
	{Bernd Ludwig}\\
	Universität Regensburg\\
	Universitätsstraße 31\\
	Regensburg, Germany \\
	\texttt{bernd.ludwig@ur.de} 
}

\renewcommand{\shorttitle}{Understanding cooking-related information needs in conversational search}

\hypersetup{
pdftitle={What can I cook with these ingredients? - Understanding cooking-related info. needs in conversational search},
pdfsubject={q-bio.NC, q-bio.QM},
pdfauthor={Alexander Frummet, David Elsweiler, Bernd Ludwig},
pdfkeywords={conversational search, information needs, conversational agents},
}

\begin{document}
\maketitle

\begin{abstract}
	 As conversational search becomes more pervasive, it becomes increasingly important to understand the user's underlying information needs when they converse with such systems in diverse domains. We conduct an in-situ study to understand information needs arising in a home cooking context as well as how they are verbally communicated to an assistant. A human experimenter plays this role in our study. Based on the transcriptions of utterances, we derive a detailed hierarchical taxonomy of diverse information needs occurring in this context, which require different levels of assistance to be solved. The taxonomy shows that needs can be communicated through different linguistic means and require different amounts of context to be understood. In a second contribution we perform classification experiments to determine the feasibility of predicting the type of information need a user has during a dialogue using the turn provided. For this multi-label classification problem, we achieve average F1 measures of 40\% using BERT-based models.  We demonstrate with examples, which types of need are difficult to predict and show why, concluding that models need to include more context information in order to improve both information need classification and assistance to make such systems usable.
\end{abstract}


\section{Introduction}
Voice-based interaction systems are changing the way people
seek information by making search more conversational \cite{guy2016searching,Trippas2017}.
Spoken queries are known to be very different to typed queries \cite{guy2016searching}. They are longer and use richer semantics and more natural language phrases compared to text queries \cite{guy2016searching}. These qualities make it possible for users to express more complex and abstract concepts than are typically used in search. Human-to-human conversation, nevertheless, goes further, typically featuring follow-up and clarification questions that lead to better shared understanding of needs and perspectives. If systems could replicate this process, conversing as humans do, systems could move beyond simply providing results to offering intelligent assistance \cite{kiseleva2016predicting}. 

Understanding how this process works in human-to-human conversation is a pre-requisite to implementing this kind of functionality into conversational assistants. Despite fields, such as Information Science, having a rich tradition of studying conversations with librarians to learn about user information needs (see e.g. \cite{Taylor1962}, \cite{Taylor1968}, \cite{French1990}) and their behaviour to address these (see e.g. \cite{belkin1987knowledge}, \cite{Boose1993}, \cite{Shute1993}), voice-based applications on mobile devices and dedicated hardware, such as
Amazon Echo and Google Home mean that information-seeking conversations now take place in diverse situations embedded in users' everyday lives that are very different to a library visit. Our understanding of how information-seeking conversations unfold is very limited in domains beyond the library. 
Recent research meetings focusing on conversational search have identified the need to accurately elicit information needs, correct user misconceptions and provide the right amount of information at the right time across all possible domains (e.g. \cite{Culpepper2018}, \cite{ananddagstuhl2019}). Our focus, in this work, is on the first of these challenges -- need elicitation -- specifically on understanding and predicting user information needs, which are important for systems to conversationally identify what a user requires, facilitate appropriate retrieval and attain relevance feedback \cite{Trippas2018}. Although domain independent taxonomies already exist, describing, for example, conversational moves for search \cite{Trippas2020} or conversations for domain agnostic planning tasks (e.g. \cite{jekat1995verbmobil}, \cite{Stolcke2000}), no domain specific schemata have been published to date. 

Prominent researchers in this area emphasise that developing new information need taxonomies derived from studies in diverse realistic settings is crucial to developing usable real-world systems (e.g. \cite{Trippas2020}, \cite{Shiga2017}). With this in mind, we study information needs in one particular domain, the domain of home cooking, which, based on the literature, we believe provides a fertile context for the kinds of complex needs suited to conversational search \cite{nouri2020,elsweiler2017exploiting,CunninghamBainbridge2013}. Cooking, moreover, offers a situation where users simultaneously perform practical, sometimes cognitively challenging tasks, which make searching in the traditional sense problematic. By studying conversational information-seeking behaviour in this setting, we can learn not only about the diverse information needs that occur, but also how information needs are communicated and resolved in this specific scenario. Being able to predict these needs is a first yet crucial step towards a truly conversational search system.

Concretely our contributions are the following:
\begin{itemize}

\item we perform an in-situ study that facilitates a naturalistic cooking situation resulting in the organic development of information needs,
\item we analyse the collected data qualitatively to learn about the diverse types of information needs which can occur in this context,
\item we utilise machine learning approaches to classify needs using the raw transcription of participant utterances. 
\end{itemize}

In doing so, our findings add to the information-seeking literature where information needs have been studied in many contexts for many reasons, but rarely in the context of home cooking. Moreover the results are insightful for the future development of conversational search systems as they show that within this context it is possible to detect the kind of need a user has based on the raw speech utterances. For systems of practical use to be developed, however, the performance needs to be improved. Our results suggest this may be achieved by providing algorithms with more context information.

\section{Related Work}
\label{sec:related-work}

Our research relates to research contributions across diverse fields of the computer and information sciences, ranging from information seeking to natural language understanding to artificial intelligence. Here we link the fields by first reviewing appropriate and relevant literature on conversations in interactive information retrieval, and understanding and predicting user needs and goals, before relating this research to appropriate contributions on natural language understanding, more generally.

\subsection{Conversations in Interactive Information Retrieval} 
\label{conversations_in_IIR}

Studying conversations is not new in information seeking. Many of the key models, which have guided research for decades, were conceived by analyzing user conversations e.g. \cite{kuhlthau1991inside,belkin1987knowledge,Taylor1962}. Belkin’s research used dialogues to model information-seeking behaviour and later dialogue structure inspired the design of interactive IR systems \cite{belkin2001iterative,yuan2014applying}. Prekop \cite{Prekop2002} identified patterns describing prototypical behaviours, actions and interactions of participants in collaborative tasks. The patterns were \enquote{Information Seeking by Recommendation} [p. 543], \enquote{Direct questioning} [p. 544] and \enquote{Advertising information paths} [p. 545]. Other publications looked at the language used in conversations. Thomas et al. \cite{thomas2018style} distinguish several conversational styles and discovered that divergent styles between agents can be strenuous for users. One take-away from the work is that conversational systems should be capable of adapting to the user's style to reduce workload. 
Trippas \cite{Trippas2017} identified patterns resembling Prekop's when pairs of users conversed during collaborative tasks. She also found that integrating discourse in conversational search models is necessary to cope with the complexity of the information-seeking process \cite{Trippas2020}. For example, grounding, i.e. making mutual knowledge and intentions clear, is a means of relevance feedback in the conversational search domain \cite{Trippas2020} which can help reduce complexity. This strategy was also found to be important by Aliannejadi et al. \cite{aliannejadi2019} in the sense that clarifying questions led to better search results since they help to resolve complex information needs. Zamani et al. \cite{zamani2020}, who investigated which kind of clarifying questions are preferred by users, also found that such questions are typically employed in ambiguous and long natural language user interactions. This indicates that the information needs are hard to express in such situations and thus might be complex. This aligns with Ruthven's \cite{ruthven2019} findings which indicate that early stage, i.e. hard to express, information needs are expressed with more words and greater narrative.  
The work by Trippas, Aliannejadi and their colleagues shows that conversational search supports users by allowing natural dialogue to address complex goals. The underlying information needs of these goals, however, are linguistically hard to express for users. For this reason, an in-depth understanding of information needs would be beneficial as it would offer the potential for conversational search systems to adapt to user needs and behaviour.

\subsection{Understanding user information-seeking goals}
The aforementioned properties of information-seeking dialogues, like adaption of the language style \cite{thomas2018style, thomas2020}, the usage of clarifying questions \cite{aliannejadi2019, zamani2020} and grounding \cite{Trippas2020}, are important aspects that help understanding information-seeking goals. However, understanding user information needs is not a new field of investigation.

Conversational search fits with the progression in IR, from the matching of queries and textual documents to understanding and supporting user needs. Classifications of search tasks are one way of thinking about user information needs. In IIR, such classifications are numerous and often delineated based on type or characteristic \cite{Kelly2015}. Type-based examples focus on what people are trying to achieve e.g. \cite{Toms2008}. 
Characteristic-based classifications, on the other hand, delineate tasks by property, such as specificity \cite{rouet2003looking} or complexity \cite{Bystrom1995}. Such classifications are useful because they inform the design of IIR systems \cite{Wildemuth2014,Kelly2015}, help to study and model search behaviour \cite{He2017}, and provide the basis for the evaluation of search systems and user interfaces \cite{Elsweiler2007}. Classifications for specific domains including web \cite{Broder2002}, mobile  \cite{Church2014} and personal search \cite{Elsweiler2007} have been published. The closest in the literature to our work is \cite{CunninghamBainbridge2013}, which establishes categories based on 678 Google Answer Forum queries relating to cooking, identifying 17 types of information need. These, however, are not based on conversational speech data and thus are only to some extent transferable to conversational search systems. Trippas \cite{Trippas2020} suggests that domain-specific taxonomies need to be created to better understand the complexity of information needs and user information-seeking goals. Extracting these domain-specific information needs from the user can be done by mixed-initiative conversational search systems who retrieve information by user-agent conversations and help the user reveal their underlying information need \cite{Radlinski2017}. To achieve this goal technically, systems must to be able to automatically predict and detect the user's information-seeking goal, i.e. information need.

\subsection{Predicting information needs conversationally}
\label{sec:related_work_predicting_in}
Understanding and algorithmically predicting user needs can  be useful for many reasons: different results can be shown \cite{craswell2001effective}, results can be presented differently \cite{teevan2009visual} or answers can be presented directly in the results page \cite{bernstein2012direct}. Conversations with the user are one means of detecting such information needs. Automated conversational agents can provide personalisation and support users in selecting an item \cite{Thompson2004} or talking about areas of interests \cite{Spillane2017} and have been applied in scenarios such as in trauma therapy \cite{Morbini2012}. This often requires systems to 
utilise a memory function where systems can reference past statements \cite{Radlinski2017} in order to maintain an understanding of context \cite{Kopp2005, Allen2001}. 

In conversational search, preliminary work has utilised user speech utterances as a means to identify user intents. Intent classification models are also used in other conversational domains including generic agents (see e.g. \cite{khalil2019cross, liu2019reconstructing, larson2019evaluation}). Focussing on the conversational search domain, Ahmadvand \cite{ahmadvand2020} attempts to capture context with the use of topic sequence models in order to suggest topics in open domain dialogues. Qu et al. \cite{qu2019} predict generic user intents with the help of dialogue acts, while Boteanu et al. \cite{boteanu2020} focus on the prediction of latent intents in text-based product queries. In addition to predicting user intents, there have been efforts to improve information need prediction. For example, Aliannejadi et al. \cite{aliannejadi2020harnessing} found that contextually relevant utterances lie at a position in the conversation that is close to the current utterance which informs the design of the memory and context property of such systems. Ren et al. \cite{ren2020} attempt to understand complex information needs and provide appropriate answers in a conversational search/agent setting. They emphasise the importance of context, e.g. via conversational history, which they argue must be modelled to fully understand a current turn. Shiga et al. \cite{Shiga2017} classify needs along two dimensions, the first of which uses Taylor’s levels of specification \cite{Taylor1962} and the second, which delineates type based on a classification derived from the literature. They, moreover, incorporate an aspect of task hierarchy, where a main task (e.g. booking a holiday) can be viewed as consisting of sub-tasks (e.g. findings a destination, comparing flight schedules etc.). Their work shows that information need categories can be distinguished using machine learning approaches. This work is the closest to our own research in terms of motivation and approach. However, the categories of needs predicted are very high level and domain unspecific which is also the case in the study by \cite{qu2019}. One could imagine that conversations and the types of support required across domains could be quite different. If systems could identify specific need types within specific domains, conversational systems could provide much more appropriate assistance. Thus, building on Shiga et al.’s work we test similar approaches in a home cooking context.

\subsection{Understanding dialogue}
\label{sec:understand_language}
Theoretical foundations and technical solutions for conversational search are closely related to research on dialogue understanding in linguistics and artificial intelligence. In the vast literature on this topic, two complementary trends are visible: the first (and historically earlier) trend relates to discourse analysis and aims to generate computational features from all linguistic, cognitive, and (discourse) pragmatic processes that are active --- often in parallel --- while dialogue participants interact. Macagno and Bigi \cite{doi:10.1177/1461445617691702} present a recent overview that positions conversational search in the class of {\it information-seeking dialogues}. While this strand of research develops extensive linguistic models for the pragmatics of dialogues, it does not focus on developing effective algorithms for these. The second major trend does exactly this, based on recent successes of machine learning algorithms (see e.g.\ \cite{DBLP:journals/corr/SerbanLCP16,wen-etal-2017-network}): Deep neural networks are leveraged to implicitly model the current state of any active process using the hidden and recurrent layers of the employed network architecture. Dialogue systems based on these machine learning approaches are regularly evaluated in the dialogue state tracking challenge \cite{DBLP:journals/corr/abs-1901-03461} on a class of so-called {\it open domain} dialogues \cite{10.1145/3383123,10.1145/3331184.3331375} or {\it multi-domain} dialogues \cite{DBLP:journals/corr/abs-2004-08114}. Whereas algorithms for open domain dialogues aim to imitate human-to-human conversations as naturally as possible, without necessarily trying to complete a task, other single-domain dialogues can be task-based (e.g. \cite{DBLP:journals/corr/abs-2001-07526}), where the user provides rational interactions trying to complete a certain goal (e.g.\ booking a flight). In this class of dialogue systems, the task knowledge is formalized in terms of slots \cite{DBLP:journals/corr/abs-2001-07526} whose values must be extracted from natural language user input using appropriate natural language understanding (NLU) algorithms. If such problems are to be extended to multi-domain dialogues, a small number of tasks may need to be completed in parallel (i.e.\ booking a flight, reserving a room in a hotel, and retrieving a public transport connection from the airport to the hotel). In either case, machine learning based dialogue systems are trained on huge corpora of (mostly publicly available) data sets \cite{DBLP:journals/corr/abs-1901-03461}.

Such systems are trained to predict values for slots to be filled from current user input and a current latent state of the task inferred from the sequence of dialogue moves (i.e.\ classified user utterances) observed so far (see \cite{10.1145/3331184.3331375}). Since dialogue move sequences result from changes in the state of the task to be completed, the training process tends to result in models fitted strongly to the specific task. This, in turn, means that models perform poorly in contexts outside the training scenario \cite{jhunjhunwala-etal-2020-multi}.

Building a (task-oriented) conversational search system for a new domain, i.e.\ a domain for which no corpus is available remains a huge challenge despite the recent progress in machine learning. Therefore, only a few cross context systems exist due to the lack of data available and the high costs involved in annotating such data (see e.g. \cite{Wang2020RecSys}, \cite{Vanzo2019sigdial}).

\subsection{Key Differentiators}
\label{sec:key_differentiators}
The summarized related work above indicates several open issues in the literature. Below we highlight these and formulate concrete research questions that we will answer with our empirical work.

To implement working conversational search systems, it is crucial to develop taxonomies or classification schemes to understand the information needs commonly occurring in one specific domain such that systems can employ appropriate strategies to elicit the user's underlying need. As, to date, no such taxonomies exist for conversational search in the cooking domain, \textbf{RQ1} aims to answer the question of \textbf{what kind of information needs arise when cooking with a conversational assistant.}

Detecting such needs in a dialogue still remains a huge challenge for current systems due to difficulties in understanding discourse pragmatics. In \textbf{RQ2} we want to investigate \textbf{how information needs are communicated} to identify linguistic features in dialogue that raise difficulties in interpreting information needs.

Finally, systems must be capable of predicting user information needs to satisfy these accordingly. Since the previous literature has only focused on predicting broad and domain-agnostic needs, we evaluate in \textbf{RQ3} whether \textbf{cooking-related information needs can be predicted in a conversational assistance scenario}.

\section{Methodology}
\subsection{Data Collection}
We designed an experiment to collect naturalistic dialogues that involved the communication of and attempts to solve information needs. To simulate a naturalistic cooking scenario, an insitu experiment was performed, whereby each participant was gifted a box of ingredients and tasked to cook a meal that they had not cooked previously using as many of the contained ingredients as possible. Participants were allowed to supplement the supplied ingredients with the contents of their own pantry if necessary. To assist this process, participants could converse with the experimenter who would answer any questions and needs arising using any resources available to him via the Web. The experimenter provided the best answer he could and communicated this orally in a natural human fashion, which is arguably the optimal behaviour for a conversational system. The setup was equivalent for all participants who worked in their own kitchens with the experimenter mimicking a conversational search system, sitting close by with access to the Internet via his laptop. No time constraints were imposed for the task. Each experiment comprised the following six steps:
\begin{enumerate}
    \item The instructions were read to the participant.
    \item Participants signed a consent form explaining how the collected
data would be stored and used in the future.
    \item The ingredient box was provided.
    \item The recording device was tested.
    \item Participants started the cooking task and the full dialogue
between experimenter and participant was recorded.
    \item After the task, the experimenter thanked the participant and
gifted the remaining ingredients.
\end{enumerate}

To encourage a naturalistic setting, all experiments took place in each participant's home. As such, the participants were more familiar with their surroundings and materials available. Feeling comfortable in the environment is advantageous as it is known to leverage informal speech \cite{OrengoCastella2000}.

\subsection{Ingredients}
To ensure diversity in the recipes and thus also variety in the subsequent conversations, the ingredient boxes varied across participants. The ingredients mostly had a value of around 10€ and were chosen based on the guidelines by the German Nutritional Society, which suggest seven categories of ingredient are required for a balanced meal. Typically the box contained some
kind of grain or starch (e.g. potatoes or rice), a selection of vegetables, a source of protein (e.g. eggs), and at least one ingredient that was seasonal  (e.g. asparagus) or that we felt might not have been commonly cooked with by all participants. We did this in order to stimulate conversation and thus to gain as many potential utterances as possible. Participants prepared diverse meals using the ingredients. To ensure freshness, all ingredients were bought on the day they were used in an experiment.

\subsection{Participants}
Participants were recruited using a snowball sampling technique with a convenience sample providing the first group of candidates. These participants, in turn, were willing to recruit friends and relatives and so on. This method offers two advantages. First, it generates a basis for trust among the participants and the experimenter, which leads to more informal and open dialogue \cite{OrengoCastella2000}. Our impressions confirmed relaxed and natural behaviour in the experiments. Second, it allows a relatively large sample to be achieved. The only requirements for potential participants were a kitchen and Internet connection. Participants were not paid for participation but to increase response rate, the ingredients were gifted. 45 participants (22 females, $\overline{x}_{age}=24$ years, $min_{age}=19$ years, $max_{age}=71$ years, $20\%$ non-students) were tested between May 7, 2018 and June 28, 2018. 37 had never used conversational agents before, while four used either Alexa or Google Home. Asked about their cooking experience, six participants reported cooking multiple times per week or on a daily basis, 18 reported cooking seldomly or not at all and one person regarded cooking as her hobby. The remaining 20 participants stated that they cooked but not on a regular basis.

\subsection{Transcription}
\label{sec:transcription_methods}
In total, 38.75 hours of material was collected with the language spoken being German. The recorded conversations were transcribed by a trained linguist. Compared to our pilot study \cite{Frummet2019}, this new transcription represents a much more rigorous and work intensive approach providing additional detail to systems and future research, i.e., pauses and intonation.

The transcription approach was inspired by previous work dealing with comparable data i.e. naturalistic conversational search dialogues in German dialect \cite{Frummet2019}. Since many participants spoke in dialect (Bavarian German), we followed the recommendation in Dresing \& Pehl \cite{DresingPehl2015} to translate the dialectal expressions into standard German so that other German readers can fully understand what was said. This is also important for the natural language processing steps that take place afterwards, which necessitate Standard German text. Syntactic errors and the word order were, however, maintained \cite{DresingPehl2015}. Building on past work from the conversational search literature, we applied the guidelines for transcribing spoken conversational search dialogues that Trippas \cite{Trippas2017transcriptionprotocol}, adopted from \cite{mclellan2003}. These recommendations include the suggestions that transcription rules should be independent in the sense that independent transcribers or third parties can easily understand them, that the transcripts should be suitable for both human and computer use and that the transcription rules should be elegant, thus, minimising the number of rules and keep them easy to learn.

More concretely, inspired by Dresing \& Pehl \cite{DresingPehl2015} and Trippas \cite{Trippas2017transcriptionprotocol}, Braun \& Clarke \cite{braun2013} and McLellan et al. \cite{mclellan2003}, we followed the guidelines listed below:

\begin{itemize}
    \item Recorded audio files were transcribed verbatim (following the guidelines above). Every dialogue related to the cooking session was transcribed. Only private chatter and small-talk during the experiment were not transcribed due to ethical issues. These passages were marked with a (smalltalk) placeholder.
    \item Noisy, incomprehensible passages were marked with (unv.) which means \enquote{unverständlich}, the German word for incomprehensible. Other notes were also marked in rounded brackets
    \item There is no punctuation in the transcript. Instead, intonation was marked in the following way: (\textunderscore) and (;) mark a low intonation whereby (;) is not as low as (\textunderscore). (-) indicates that intonation is on the same level, i.e. neither going down nor up. (,) and (?) indicate a high intonation where (?) is higher than (,). For further investigation with the usage of intonation it is highly recommended to group the high and low intonation levels.
    \item Pauses were added with the following pattern: (.) is a short break of about one second, (..) is a break of about two seconds, (...) is a three second pause and ([number]) is a pause lasting [number] seconds.
    \item Numbers up to twelve were transcribed as words, higher numbers were transcribed numerically.
    \item URLs were transcribed as they were spoken (e.g. chefkoch.de was transcribed as chefkoch d e)
    \item Abbreviations were written as said.
    \item Overlaps in talking were marked with double slashes: //
    \item Strong accentuation was marked with capital letters.
\end{itemize}
\subsection{Annotation}
\label{sec:annotation_methods}
 
One primary goal is to achieve a detailed understanding of the information needs that occur during the cooking process, as well as how these originate and are expressed conversationally. We previously published a preliminary analysis of the collected data using less rigorous transcription and analysis approaches \cite{Frummet2019}. The results of the preliminary work were sufficient to convince us of the value of insights that could be provided using a more detailed analysis. More detail in the transcription and analysis processes are necessary as the information need categories in our pilot study were very broad and therefore difficult to distinguish using machine learning approaches. Moreover, hierarchical relationships and semantic similarity between categories lead to miss-classifications. By creating a more precise, tree-structured representation of the information needs in the cooking context, our aim was to develop more accurate classifiers. By doing this, we aim to improve the performance of classifiers when distinguishing between different types of information need. 

The process comprised the following steps outlined by Braun \& Clarke \cite{braunclarke2006}: (Step 1) familiarising self with data, (Step 2) generating initial codes, (Step 3) searching for themes, (Step 4) reviewing themes, (Step 5) defining and naming themes, and (Step 6) producing the report. Furthermore, we followed the recommendations by Charmaz \cite{charmaz2006} on line-by-line-coding which helps to remain open during the coding process and to detect more nuances in the data. One crucial element for comparison and consistency validation was the strong usage of memos where decisions taken and new ideas were noted as suggested by Corbin \cite{corbin2016}. Using the \textit{f4analyse}\footnote{see https://www.audiotranskription.de/en/f4analyse/} software made iterative comparison and refinement convenient due to support for qualitative data analysis (QDA) which allows utterances from multiple classes to be compared, making the identification of commonalities and differences easier. The whole annotation process was conducted by the first author who is also a trained linguist. As outlined by Braun and Clarke \cite{braunclarke2006}, the process was conducted iteratively and tested repeatedly until the researcher was satisfied a consistent information need taxonomy was achieved. The following steps were taken to ensure the development of a consistent coding scheme: 
\begin{enumerate}
    \item Identifying which participant spoke, i.e. identifying turns. For turns in spoken conversational search the definition by Trippas \cite{Trippas2017transcriptionprotocol} was employed: \enquote{Taking the initiative equals taking the turn.} Thus, a turn could consist of multiple moves or conversational goals.
    \item Assign codes to each turn and take notes and memos for later comparison and to understand why certain decisions have been made.
    \item Combining codes to themes for further analysis and merging indistinguishable labels.
    \item Check for consistency across participants and themes to assure that the labels actually are distinct.
\end{enumerate}

Selective coding was employed to systematically relate codes, validate the relationships between them and add and refine further categories as suggested by Strauss \& Corbin \cite{StraussCorbin1996}. By doing this, we gained the tree-like structure of our information need taxonomy. This helps us to better understand the relationships between the information needs.

\section{Results}

The results are presented in three subsections, each of which addresses one of the research questions formulated in Section \ref{sec:key_differentiators}. In Section \ref{sec:info_need_taxonomy}, we address RQ1 by presenting a hierarchical information need coding scheme along with its top-level categories. These categories are described by means of examples in Table \ref{tab:table_top_in_cat_part}. Section \ref{sec:in_communication} addresses RQ2 and illustrates how information needs were communicated linguistically by participants. This allows us to identify dialogue-specific challenges that need to be considered when implementing systems for information need detection. To address RQ3, Section \ref{sec:predicting_IN} reports on experiments designed to establish the extent to which the class of information need (based on the level 1 categories) can be successfully predicted given participant utterances. 

\subsection{Information Need Taxonomy}
\label{sec:info_need_taxonomy}
We first offer a broad overview of the different levels of information needs detected in our corpus and provide a sample description of the utterances issued to the experimenter. We subsequently present a more detailed description of the tree structure and the taxonomy codes to offer a more detailed understanding of the information needs that occur when cooking. We refer to our annotated dataset as the \textit{CookversationalSearch} dataset.
\subsubsection{The CookversationalSearch dataset}
\label{sec:sample_description}
\begin{figure}[htpb]
  \includegraphics[width=.8\linewidth]{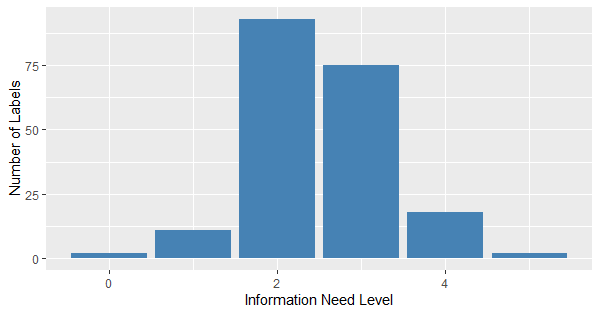}
  \caption{Information need distribution over levels.}
  \label{fig:information_need_level_dist}
\end{figure}

contains, overall, $N_{t}=2376$ turns describing one or multiple user information needs that were extracted from the dialogues\footnote{https://github.com/AlexFrummet/CookversationalSearch}. 2102 ($88.5\%$) of these express one information need only, 249 ($10.5\%$) of all turns represent two information needs and 25 ($1.05\%$) turns express three different needs. Each row in the dataset represents one turn and its corresponding information need. Thus, some turns appear multiple times, i.e., in multiple rows, as they can represent multiple needs. This results in a corpus with $N_{t}=2675$ turns in total, with each representing an information need and consisting of one or more sentences. We also refer to such turns as queries, thus the terms \textit{query} and \textit{turn} will be used interchangeably for the rest of this article. 
Each trial yielded an average of $52.80$ queries ($min=8$, $x_{.25}=33$, $\tilde{x}=45$, $x_{.75}=70$, $max=120$, $sd=28.73$, $skewness=0.57$, $curtosis=-0.60$). The coding scheme used to describe these queries is structured as a tree. The tree consists of two top level (= level 0) branches, each of which descends, at deepest, to a 5th level. Each level provides additional detail describing the user's information need. To illustrate this process we provide samples from the dataset in Table \ref{tab:coding_sample}. Readers can additionally access the full coding scheme\footnote{https://github.com/AlexFrummet/CookversationalSearch/blob/master/annotation/InfoNeedTaxonomy.svg}. Turns were coded at the lowest level possible. 
 Figure \ref{fig:information_need_level_dist} shows the number of codes distributed over the different information need levels. It can be seen that in addition to the two level 0 categories, 11 information need categories occurred on level 1, 93 information needs occurred on level 2, 75 on level 3, 18 on level 4 and 2 on level 5. All 2376 turns were assigned a level 0, level 1 and level 2 code. Thus,  all turns were coded to at least level 2. 1848 queries were assigned with a level 3 code, 395 with level 4 and 97 with level 5. More details about the different levels and information needs will be provided in the following sections.
\subsubsection{Inter-rater reliability} 

To check the consistency of the annotation scheme, inter-rater reliability with a second researcher who is also an experienced coder was measured. Before calculating agreement, small samples from the corpus were coded by both annotators in multiple iterations. Problems were discussed and the codebook was refined after each round. Utterances representing an information need were labelled with annotators having access to full dialogues and context to better interpret an utterance. In the final pass, a stratified sample of 10\% of all corpus utterances was selected among level 1 categories. Stratifying the sample allowed us to test whether agreement could be achieved for all information need categories.  

To measure inter-rater agreement, unweighted Cohen's $\kappa$ was chosen. Artstein \& Poesio \cite{artsteinpoesio2008survey} suggest this measure for dialogue data where category labels are equally distinct from one another. In the final pass, the annotators achieved a Cohen's $\kappa$ value of $\kappa$=0.69 using codes at level 2 (93 codes) and a value of $\kappa$=0.75 using codes at level 1 (11 codes), which according to Landis \& Koch \cite{landiskoch1977kappa}, both represent substantial agreement. Although there are many more codes on level 2 than on level 1, this did not relate to a substantially lower kappa value.

The agreement achieved is higher than what has been achieved in other conversational search corpora that were created in much more controlled contexts. For example, when evaluating the SCoSAS annotation schema, a kappa score of $\kappa$=0.59 was achieved \cite{Trippas2020}. Moreover, on the widely used Maptask dialogue corpus -  which was collected in a very controlled situation - the same level of inter-rater reliability ($\kappa$=0.59) was attained according to Isard \& Carletta \cite{isardcarletta1995}.
\begin{table}[htpb]
\centering
\begin{tabularx}{\textwidth}{XX}
\toprule
\textbf{Information Need} & \textbf{Cohen's $\kappa$}              \\
\midrule
Amount            & 0.96 \\
Cooking technique & 0.72 \\
Equipment         & 0.89 \\
Ingredient        & 0.59 \\
Knowledge         & 0.66 \\
Meal              & 0.67 \\
Misc              & 0.86 \\
Preparation       & 0.62 \\
Recipe            & 0.90 \\
Temperature       & 1.00 \\
Time              & 0.97 \\
\bottomrule
\end{tabularx}
\caption{Cohen's $\kappa$ values for each level 1 information need category.}
\label{tab:inter_rater_agreement}
\end{table}

\renewcommand{\arraystretch}{3}
\newcolumntype{b}{X}
\newcolumntype{s}{>{\hsize=.5\hsize}X}
\newcolumntype{l}{>{\hsize=.2\hsize}X}
\begin{table}[htpb]
\begin{adjustbox}{angle=90}
\small
    \begin{tabularx}{\textheight}{sssbb}
    Level 1 Code &  Level 0 Code & Frequency & Definition & Examples\\
    \midrule
    Preparation & Competence & 768 ($28.71\%$) & Utterances were labelled with this category when queries related to a particular step of the recipe (as opposed to general cooking techniques, see below). & \enquote{What's next after bringing to the boil?}, \enquote{What should I do after the potatoes are prepared?}, \enquote{Ok, that means that I now have to add the flour, right?} \\
    Ingredient & Fact & 658 ($24.60\%$) & Whenever questions regarding which ingredients were necessary for a particular recipe occurred, these utterances were tagged with label \textit{Ingredient}. & \enquote{Which ingredients are needed?}, \enquote{Please read the ingredients aloud.}, \enquote{What [ingredients] goes in a vegetarian casserole?} \\
    Amount & Fact & 476 ($17.79\%$) & The label \textit{Amount} was used, to code queries from participants who wanted to know about the quantity of an ingredient needed. & \enquote{How much egg yolk is needed?}, \enquote{How much of the couscous do I need for one person?}, \enquote{150 grams of tomatoes?} \\
    Cooking Technique & Competence & 210 ($7.85\%$) & Utterances/Queries were labelled this way when participants requested information about preparing ingredients that was not made explicit in the steps of the recipe. & \enquote{But it has become a bit too watery. It would be good if\dots How can I reduce the liquid?}, and \enquote{Ok how do you prepare potatoes properly?}\\
    Recipe & Fact & 205 ($7.66\%$) & \textit{Recipe} was used in cases where participants searched for recipes they would like to prepare as their main dish. They often used ingredients as search items in such cases. & \enquote{Then I'd like to have a dish with lentils, chickpeas and tomatoes}, \enquote{Ok, search for a recipe with couscous and vegetables.}, \enquote{Is there a recipe for Jerusalem-Artischocke and celery soup?}\\
    Time & Fact & 182 ($6.80\%$) & Any utterances regarding cooking time needed were labelled with the category time. & \enquote{Ok, how long does it take? 10 minutes or 20 minutes?}, \enquote{Er - how long do I need to bubble the, the, the lentils?}
\end{tabularx}
\end{adjustbox}

\end{table}
\begin{table}[htpb]
\begin{adjustbox}{angle=90}
\small
    \begin{tabularx}{\textheight}{sssbb}
    Level 1 Code & Level 0 Code & Frequency & Definition & Examples\\
    \midrule
    Knowledge & Fact & 49 ($1.83\%$) & \textit{Knowledge} was applied to all queries/questions relating to definitions. It was also assigned to general knowledge questions, i.e. questions that did not relate to the implementation of a cooking step but were of more general nature. & \enquote{Can lemons go bad?}, \enquote{Teaspoons are the small ones, right?}, \enquote{What happens to sour cream when it is heated?} \\
    Miscellaneous & Fact & 44 ($1.64\%$) & Needs and commands that didn't fit into any of the other information need classes and occurred merely rare were gathered here. & \enquote{Alex, name some good pizza delivery services.}, \enquote{Am I allowed to modify the recipe?}, \enquote{Ok - er- do you have an integrated timer?}\\
    Equipment & Fact & 40 ($1.50\%$) & Utterances were labelled with \textit{Equipment} when queries/questions regarding the cooking utensils needed occurred. & \enquote{Can I use a pot for this?}, \enquote{How big should the pot be? Small, medium, large?}\\
    Temperature & Fact & 31 ($1.16\%$) & Any utterances regarding the heating type and the temperature at which ingredients and meals should be cooked received this label. & \enquote{Do I need to preheat the oven?}, \enquote{At which temperature?}, \enquote{I need to heat that, right?} \\
     Meal & Fact &12 ($0.45\%$) & \textit{Meal} includes information needs that are related to the current meal, but from a  high-level perspective. This category does not include information needs that are particularly related to the preparation steps or the ingredients used in the recipe. While the recipe class is focused on the process of recipe selection, this class includes turns after a recipe has been chosen. & \enquote{Which drinks go with chili?}, \enquote{Can this be compared to stew?}
    \end{tabularx}
\end{adjustbox}
\caption{Level 1 information need categories with their corresponding level 0 code.}
\label{tab:table_top_in_cat_part}
\end{table}
\renewcommand{\arraystretch}{1.0}
\begin{table}[htpb]
\small
    \begin{tabularx}{\linewidth}{bssssss}
    \toprule
    \textbf{Turn} & \textbf{Level 0} & \textbf{Level 1} & \textbf{Level 2} & \textbf{Level 3} & \textbf{Level 4} & \textbf{Level 5} \\
    \toprule
    \enquote{Um can you find me dishes with asparagus with many dairy products.} (part. 1) & Fact & Recipe & Recipe Retrieval & Recipe Request& Recipe Request with Ingredients & Explicit \\ \midrule
    \enquote{How much sugar?} (part. 2) & Fact & Amount & Amount of Ingredients & -- & -- & --\\ \midrule
     \enquote{Um -- How do you prepare bulgur?} (part. 34) & Competence & Cooking technique & Cooking technique - Ingredient & -- & -- & -- \\ \midrule
    \enquote{Then I need to add the onion, right?} (part. 40) & Competence & Preparation & Preparation Steps & Preparation Step - Explicit Reassurance & -- & -- \\
    \bottomrule
    \end{tabularx}
    \caption{Coding samples from the dataset.}
    \label{tab:coding_sample}
\end{table}

\subsubsection{Description of the communicated Information Needs}
\label{sec:in_description}

\begin{figure}[htpb]
  \includegraphics[width=\linewidth]{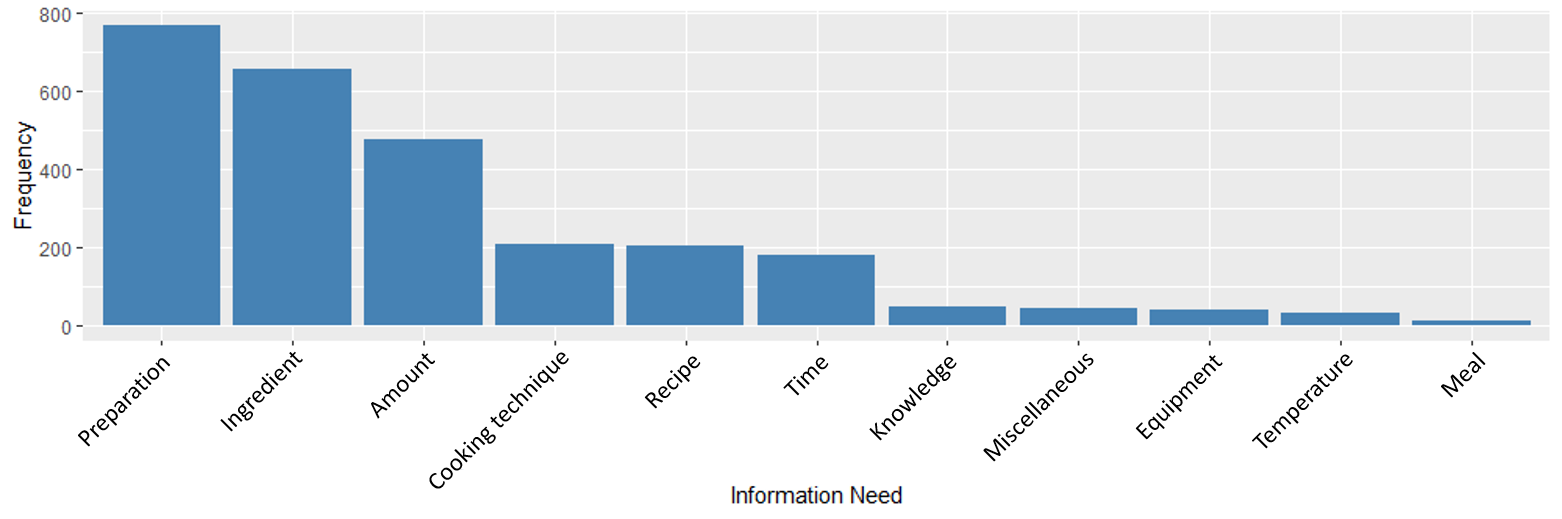}
  \caption{Level 1 Information need frequencies.}
  \label{fig:information_need_freq}
\end{figure}

This section provides a broad outline of the information needs as derived from the qualitative analysis. For space reasons, we limit our descriptions to the two top-most levels, i.e. the two information need types on level 0 and the 11 information need categories on level 1\footnote{The codebook with a detailed description of every information need class is available online: https://github.com/AlexFrummet/CookversationalSearch/.}. 
Level 0 distinguishes between \textit{Fact-based} needs which relate to information about recipes and \textit{competence} needs which refer to knowledge about how to prepare recipes. Level 1 delineates needs more precisely, and includes codes, such as \textit{Recipe} which relates to the user seeking a recipe to cook, and \textit{Ingredient}, which relates, in some way, to ingredients that are contained within the recipe being cooked. To assist the reader’s understanding, we provide several examples of queries for each class of information need at level 0 and 1 in Table \ref{tab:table_top_in_cat_part}. Figure \ref{fig:information_need_freq} shows the frequency of queries over the different types of information needs. Below, we first illustrate the hierarchical annotation scheme with a short example. We then describe the defining characteristics of different information need types in greater detail.

\paragraph{Hierarchical annotation:}
\label{sec:hierarchical_annotation}
As described in the previous section, the annotation scheme has a tree-like structure. A visual representation of this taxonomy can be found in the accompanying repository\footnote{https://github.com/AlexFrummet/CookversationalSearch/blob/master/annotation/InfoNeedTaxonomy.svg}. To illustrate the hierarchical annotation process, we provide some examples from the dataset in Table \ref{tab:coding_sample}. Lower-level codes subsume the code from the higher-level branches. The turn \enquote{How much sugar?} (part. 2) from Table \ref{tab:coding_sample}, for instance, is coded as \textit{Amount of Ingredients} on the lowest level and, thus, inherits the \textit{Amount} category from level 1 and \textit{Fact} from level 0.


\paragraph{Fact-based Information Needs}
\label{sec:fact_based_in}
refer to information about the recipe being cooked.
One prototypical fact-based information need is the level 1 category \textit{Amount}, which includes turns that target the quantity of e.g. ingredients recipe portions. Information needs of this type are easily identificable due to their specific vocabulary and syntax. The majority of queries that were labelled with \textit{Amount} start with keywords like \enquote{How many...} or \enquote{How much...} as in \enquote{How many onions do I need?} (part. 25). This makes it straightforward for the annotator to label such utterances correctly. The inter-rater agreement of $\kappa$=0.96 for this category is much higher than the average $\kappa$=0.75 over all categories. It is similarly straightforward to annotate the information need categories \textit{Equipment}, \textit{Ingredient}, \textit{Temperature} and \textit{Time}. \textit{Time} for example contains many phrases, such as \enquote{How much longer does the bulgur need?} (part. 25), and \enquote{How many minutes are remaining?} (part. 29). Time-related words, such as \enquote{minutes}, and time-phrases, such as \enquote{How long...}, occur frequently which makes it easy to code utterances as time related information needs. This impression is evidenced by the relatively high inter-rater agreement in the Time category ($\kappa$=0.97). 

Keywords and phrases which are specific for certain information needs made it easier for the coder to identify needs. One commonality of fact-based needs is that little context is needed to identify this kind of information need in the dialogue.

\paragraph{Competence oriented Information Needs, }
\label{sec:competence_in}
in contrast, were more challenging to annotate because context was often required to decide the appropriate label. Competence oriented information needs are needs where the user requests information from the system on how to perform certain cooking actions. Categories associated with this type of need are \textit{Preparation} and \textit{Cooking technique}. Turns in the \textit{Preparation} category relate to particular steps of the recipe being cooked. In such cases, a user needs assistance with respect to what to do next and how to do this. When turns occurred that relate to more basic cooking processes - often not described in the recipe - 
the system (i.e. the human mimicking a system) made use of a search engine to provide assistance. If this was the case, then utterances were labelled as \textit{Cooking technique. 
For utterances in the Preparation category, the user is looking for information regarding the steps to perform, e.g. what they should do next.} Being able to differentiate between these two categories requires an understanding of the context in which a query was issued. As an annotator, one needs to understand the current task state of the cooking process and know which steps have already been completed. In contrast to fact-based information needs, the task state needs to be considered, which makes it more challenging to correctly label these kind of needs. Unlike fact-based needs, it was hardly possible to distinguish competence needs based on keywords alone. This can be illustrated by the examples \enquote{Can I add them before the water boils?} (part. 12, labelled as \textit{Cooking technique}) and \enquote{So I need to let the water boil first?} (part. 10, labelled as \textit{Preparation}). The first query was related to a more general cooking process and was, thus, labelled as \textit{Cooking technique} while the second query was concerned with the order of steps, and was therefore coded as \textit{Preparation}. It was only after the context of situation was considered that this distinction could be made. The difficulties that arise when coding such needs is reflected in the achieved $\kappa$ values ($\kappa=0.72$ for Cooking technique and $\kappa=0.62$ for Preparation) which are below the average that was achieved for fact-based needs.
~\\
~\\
To further investigate why some information need categories were easier to detect than others, we performed a keyword analysis to assess the importance of certain words for different information need categories. To measure this, we employed log-likelihood analysis using odds-ratio as effect-size measure and bonferroni adjusted p-values as statistic threshold. The results for the words with highest log-likelihood values among several information need categories are illustrated in Table \ref{tab:top15_words}. The results highlight that fact-based information needs \textit{Amount, Time and Temperature} have many more significant keywords than competence oriented information needs \textit{Cooking technique} and \textit{Preparation}. While the most distinctive words for Cooking technique and Preparation are stopwords like \textit{you} and \textit{then}, these kind of words are not found to be discriminatory in Amount, Time and Temperature. The most relevant words for Amount are \textit{how}, \textit{much} and \textit{gram} which is plausible for this information need. Similarly, it is plausible that \textit{long} and \textit{minutes} are especially relevant for Time and \textit{heat} is an important keyword for Temperature. These observations corroborate our impression that fact-based needs are easier to distinguish based on keywords than competence needs.
\renewcommand{\arraystretch}{1.0}
\begin{table}[htpb]
    \centering
    \begin{tabularx}{\linewidth}{sssss}
        \toprule
        \textbf{Amount} & \textbf{Time} & \textbf{Temperature} & \textbf{Cooking Technique} & \textbf{Preparation}\\
        \toprule
        wie \st{how}* & lange \st{long}* & hitze \st{heat}* & man \st{one}* & dann \st{then}* \\
        viel \st{much}* & minuten \st{minutes}* & auf \st{on}* & spargel \st{asparagus}* & schritt \st{step}* \\
        gramm \st{gram}* & wie \st{how}* & mittlerer \st{middle}* & schälen \st{peel}* & jetzt \st{now}* \\
        zwei \st{two}* & lang \st{long}* & stufe \st{level}* & am \st{at}* & und \st{and}* \\
        liter \st{liter}* & kochen \st{cook}* &  heiß \st{hot}* & ob \st{if} & weiter \st{further} \\
        esslöffel \st{tbspn}* & zehn \st{ten}* &  welcher \st{which}* & kann \st{can} & erst \st{first} \\
        milliliter \st{milliliter}* & dauert \st{takes}* & ofen \st{oven}* & besten \st{best} & muss \st{need} \\
        hälfte \st{half}* & bis \st{to}* & niedrigster \st{lowest}* & kocht \st{cook} & die \st{the}\\
        personen \st{persons}* & braucht \st{need}* & ober \st{upper}* & wasser \st{water} & nächstes \st{next}\\
        eine \st{one}* & uhrzeit \st{time}* & umluft \st{circul. air}* & grünen \st{green} & machen \st{do}\\
        wasser \st{water}* & zwanzig \st{twenty}* & unterhitze \st{b. heat}* & schneidet \st{cut} & fertig \st{ready}\\
        viele \st{many}* & fünf \st{five}* & vorheizen \st{preheat}* & mach \st{do} & nächste \st{next}\\
        für \st{for}* & ungefähr \st{about}* & grad \st{degree} & weißen \st{white} & zuerst \st{first}\\
        halber \st{half}* & zwölf \st{twelve} &  mittlere \st{middle} & zu \st{to} & als \st{as}\\
        menge \st{amount}* & zeit \st{time} & runter \st{down} & abschälen \st{peel} & der \st{the} \\
        \bottomrule
    \end{tabularx}
    \caption{Keywords of level 1 information need categories Amount, Time, Temperature, Cooking Technique and Preparation (arranged in descending order along log-likelihood values), with * : $p<0.05$ (bonferroni adjusted).}
    \label{tab:top15_words}
\end{table}
~\\
~\\
Section \ref{sec:info_need_taxonomy} has presented a robustly coded taxonomy with different levels of information needs. At the highest level information needs can be broadly divided into two types -- fact- and competence oriented -- which require different types of assistance from the system and different amounts of context information to be interpreted correctly.
\subsection{Communication of Information Needs}
\label{sec:in_communication}
A second research aim was to investigate how information needs are communicated. An understanding of how users verbally pose queries to conversational systems would allow us to better extract information needs from these turns by focusing on certain linguistic features. We can derive implications for the system design by knowing which linguistic aspects in the dialogue make detection and interpretation of turns as particular types of information need difficult.
Since the speech used in this corpus is natural and communicated as humans typically do, this might help us to learn about what features are relevant for information need detection for humans to make these also available for system-side information need detection. Following (stylistic) peculiarities of issuing an information need could be observed from the data:

\subsubsection{Explicit versus implicit queries}
\begin{figure}[htpb]
\centering
  \includegraphics[width=.5\linewidth]{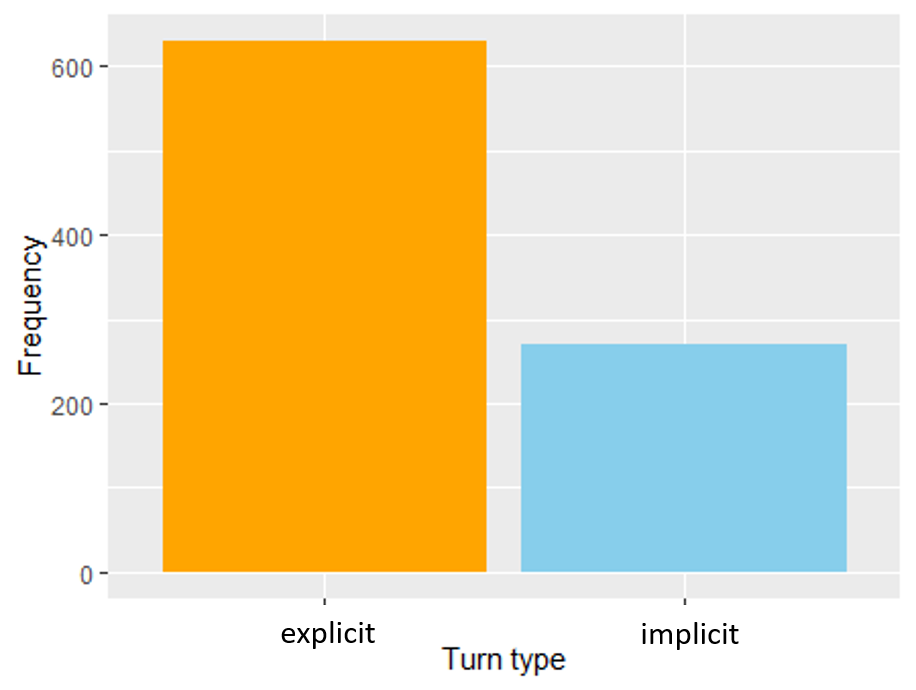}
  \caption{Distribution of implicit and explicit coded requests over all classes.}
  \label{fig:implicit_explicit_distribution}
\end{figure}

We observed both explicitly and implicitly expressed queries. Explicit queries took the form of a direct question, including an increase in intonation when the query was uttered. Explicit queries also exhibited the grammatical form of questions e.g. \enquote{Which ingredients are needed?} (part. 1). 
Implicit queries, on the other hand, largely do not possess an interrogative particle and the speaker tends to go down in intonation, as would be expected in a typical declarative sentence. An example from the corpus is \enquote{Ok. And I forgot the other ingredients.} (part. 9).
The difference between these two types of queries was so pronounced that a distinction in the annotation schema was made to reflect this \textit{Ingredients in Recipe - implicit} and \textit{Ingredients in Recipe - explicit}. Figure \ref{fig:implicit_explicit_distribution} illustrates the distribution of explicit and implicit labels. It should be noted that this distinction was only made when it was obvious. Otherwise, explicit queries did not receive a separate code. This is also the reason why Figure \ref{fig:implicit_explicit_distribution} does not split all queries into one of these two categories. What it does show, however, is that explicit queries are more frequent than implicit ones. Implicit queries, nevertheless, represent a significant portion of utterances in the corpus and thus need to be further analysed to identify patterns which help identify them in a dialogue. As outlined in Section \ref{sec:understand_language}, conversational agents often only analyse the surface structure of utterances, which makes it hard to fully understand dialogue pragmatics. The human annotators found it harder to identify implicit queries  due to their declarative and indirect shape compared to a direct question. For example, \enquote{I will just cut it in pieces.} (part. 8)
illustrates this challenge. In this case, the \textit{Preparation Step - Implicit Reassurance} label was assigned to this turn. Some people might not interpret this as an implicit query, but rather as an example of simply thinking-aloud. However, this utterance is embedded in a reading situation of preparation steps and was thus interpreted as an implicit query to reassure the next action. This illustrates that previous turns, i.e., contextual information, needs to be taken into account, especially when implicit queries should be detected. 

As already mentioned, using both explicit and implicit ways of expressing queries was a common pattern observed in the corpus. Thus, dealing with both explicit and implicit queries should be taken into consideration for systems design and will be further discussed in Section \ref{sec:implication_conv_ass}.

\subsubsection{Reassurances}
A further means that participants used to communicate information needs was to make use of reassurances. These were used extensively by participants to discover if they had correctly interpreted the instruction provided by the assistant. Linguistically, these are marked either with explicit queries, such as \enquote{Meanwhile, I have to let it cook, right?} (part. 1)
or with implicit queries like \enquote{Er - steeping it.} (part. 2). 
Implicit queries offer more scope for interpretation which made it difficult to determine whether something should be treated as query or not, as already discussed in the section before. When using reassurances in this way, participants expected an appropriate response from the assistant. Thus, this is an interaction pattern which should be considered for system design. 
To present a satisfying response to the participant, the assistant had to track the task state to inform the participant if the currently performed task is correct or not. Here, again, context information is an important feature for response generation.

\subsubsection{Discourse markers and linguistic feedback}
\label{discours_marker}
Discourse markers, such as \enquote{Mhm} (part. 1) and \enquote{Ähm} (part. 26) were frequently used in the cooking experiments. This was particularly noticeable in the classes \textit{Next ingredient} and \textit{Next preparation step}. Discourse markers relate segments pragmatically \cite{fraser1999dm} and in our corpus they were mostly used to signal that the system instruction had been understood and it is appropriate for the assistant to move to the next step in the process.  In general, in this corpus a lot of discourse markers occur since they pragmatically link conversational units in spoken language \cite{bussmann2008}. The following example from the corpus illustrates their usage:

\enquote{Participant: Ok. Good. (Alarm bell rings). Now we need to add zucchini and corn, right? Zucchini, corn and -

Assistant: Exactly. You need to mix corn, zucchini and chickpeas.

Participant: \textbf{Mhm?}

Assistant: Fill it up with half a liter of water and add the tomato paste.} (part. 5)

This example shows that the discourse marker, highlighted in bold, signals the assistant to read out the next step in the recipe. The example also demonstrates that discourse markers require context and cannot be interpreted in isolation. \enquote{Mhm?} in the example can only be interpreted as \textit{Next Preparation Step} if the content of previous turns, i.e. the reassurance question of whether mixing zucchini and corn is correct, is known. Deriving this information from the context was quite simple during annotation since this is a common situation in everyday dialogues, where Discourse Markers like \textit{Mhm?} are used as linguistic feedback \cite{allwood1992}. Its basic functions include an understanding of the currently heard and the reaction of the conversational partner to go on to the next point in the conversational flow \cite{allwood1992}. This stylistic property should be taken into account in future systems design as it frequently occurred in this corpus and brings structure into conversation \cite{schiffrin_1987}. For this reason, it is a relevant linguistic aspect to deal with as a system to better understand natural language.
~\\
~\\
This section has illustrated how three important linguistic features shaped conversational descriptions of needs. The use of explicit and implicit queries, reassurances as a means to request feedback from the assistant and discourse markers all featured heavily in user behaviour.
~\\
~\\
So far, we have analysed which information needs can occur in cooking conversational search (see Section \ref{sec:info_need_taxonomy}) and how these are communicated throughout the dialogue (see Section \ref{sec:in_communication}). One theme that is relevant for both information need understanding and detection is \textbf{context}. It is important for understanding competence needs, but is also highly relevant for identifying implicit queries, reassurances and discourse markers in a dialogue. Since context helps humans for interpretation of dialogue and needs, in the following section we want to test if systems are able to utilise context information to predict information needs automatically.
\subsection{Predicting Information Needs}
\label{sec:predicting_IN}
Now that we understand more about what information needs occur in cooking situations and how these are expressed, in this section we explore whether the turns contain enough information for systems to determine the kind of information need.
To test this, we formulate the problem as a multi-label prediction task since a single turn may express one or more information needs: Given a set of features derived from the raw conversational turns, is it possible to predict the level 1 category of information need?
\subsubsection{Dataset}
For the classification task, the \textit{CookversationalSearch} dataset which has been described above was employed. 
In our quantitative analysis, all $N_{t}=2675$ turns with their corresponding information need label were used. As outlined in Section \ref{sec:sample_description}, each turn consists of one or multiple sentences. 
On average, one level 1 category contains $243$ turns ($min=12$, $x_{.25}=42$, $\tilde{x}=182$, $x_{.75}=343$, $max=768$, $sd=269.29$, $skewness=0.85$, $curtosis=-0.96$).
\subsubsection{Experimental conditions}
The multi-label classification problem is treated as a binary classification task whereby we attempt to classify needs at level 1 in the taxonomy. Each target turn is classified as either an information need type or not, i.e. 11 classifiers (\textit{[information need]} vs. \textit{Other}) were trained -- one for each need type. Since we found context to be an important factor for identifying information needs throughout the coding process and in the literature (see \cite{aliannejadi2019, ren2020, Trippas2020}, we tested following conditions which model context information in the following ways: 
\begin{description}
\item[No context] --  We only consider the target turn, ignoring what has been uttered previously.
\item [One previous participant turn] -- We add the previous information need turn that was issued by the participant to the target turn. The previous turn and the target turn are simply concatenated.
\item [All previous participant turns] -- We include all participant turns preceding the target turn. The collected turns are added as text before the target.
\end{description}
We only account for the turns uttered by the participant, since this has been shown to be very effective in similar prediction tasks, e.g. \cite{fornaciari2021bertective}.
To avoid overfitting, we employed 10-fold cross-validation. Due to the highly skewed and imbalanced dataset, we further used stratified sampling to ensure a fair distribution in training and test sets. We also tested hierarchical classification methods, but found these to significantly lower prediction performance and thus do not report this here. The observation that hierarchical classifiers lower prediction performance has also been made by Fornaciari \& Poesio \cite{fornaciari2021bertective}.
\subsubsection{Models}
\paragraph{Baseline Models:}
To generate a baseline for our experiments, we employed some of the models, which have been shown to be effective in past experiments predicting information needs in conversational settings \cite{Frummet2019, schwabl2021classifying, Shiga2017}. 
For our baseline models, we used 300-dimensional FastText embeddings \cite{joulin2016fasttext} combined with BytePairEmbeddings provided by Flair \cite{akbik2019flair}, which are able to capture out of vocabulary words using sub-word information \cite{heinzerling-strube-2018-bpemb}. This feature was the main reason for choosing FastText embeddings as it was considered helpful when dealing with spoken dialogue data where out of vocabulary words might occur more likely due to its natural style. For the baseline experiments, we use Gaussian Naive Bayes (=GNB or GaussianNB), Random Forest and SVM classificators provided by the Python package \textit{scikit-learn} \cite{Pedregosa2011}.
\paragraph{BERT-based models:}
Since Schwabl \cite{schwabl2021classifying} showed that information need detection in the cooking domain benefits from using BERT-based models, we employed these for comparison. BERT was chosen due to its success in natural language understanding problems \cite{wolf2019huggingfaces}, such as Named Entity Recognition and Text Classification. For our experiments, we leveraged pre-trained BERT models, which are GermanBERT\footnote{taken from https://huggingface.co/bert-base-german-cased} and the multilingual BERT models mBERT \footnote{taken from https://huggingface.co/bert-base-multilingual-cased} and XLM-RoBERTa\footnote{taken from https://huggingface.co/xlm-roberta-base}. Due to the small amount of data, 85\% was kept as training and 15\% as test data to ensure that there are still samples available for the test dataset even for categories with few data. For each BERT model, we chose the $BERT_{BASE}$ architecture which was available for all three models in contrast to $BERT_{LARGE}$. Since fine-tuning pre-trained BERT models on downstream tasks has shown to improve classification performance \cite{sun2019finetune}, this technique was also applied in our experiments. All BERT related experiments were conducted using the transfer-learning library FARM \footnote{see https://github.com/deepset-ai/FARM}
To fine-tune each of the 11 binary classifiers, a TextClassficationHead \footnote{see https://farm.deepset.ai/api/modeling.html\#farm.modeling.prediction\_head.TextClassificationHead} was added on top of the BERT language model. The TextClassificationHead consists of a one layer feed-forward neural network (FFNN) of dimension (768,2) and a softmax activation function. This follows the recommendations made by Sun et al. \cite{sun2019finetune} and Devlin et al.\cite{devlin2019bert}. The classification head receives the final hidden state of the [CLS] token which represents the turn. Stopwords were not removed since removing stopwords decreases the effectiveness of the BERT model \cite{DaiCallan19}. Also, resampling was not conducted as BERT handles imbalanced datasets well \cite{madabushi2019costbert}. Class weights were adjusted by FARM's DataSilo which also ensured stratified sampling for cross validation\footnote{see https://farm.deepset.ai/api/data\_handling.html\#farm.data\_handler.data\_silo.DataSiloForCrossVal}.
We followed the suggestions by Sun et al. \cite{sun2019finetune} and Devlin et al. \cite{devlin2019bert} regarding the parameters to chose for fine-tuning. To avoid the ``catastrophic forgetting problem'', we selected a lower learning rate (2e-5) \cite{sun2019finetune}. The classifiers were trained for 4 epochs with a dropout probability of 10\% and a batch size of 32. To avoid overfitting, early stopping was included. The models were trained on three GeForce GTX 1080 Ti GPUs. Due to computing limitations, we had to choose a maximum sequence length of 256 which would have resulted in clipping down 63\% of all samples to 256 tokens for the \textit{all prev turns} condition. As preceding turns were added as raw text before each target turn in this condition, leaving the text unchanged would have led to cutting out the actual target turn. Thus, as a preprocessing step, only the last 256 tokens which also include the current target turn were considered. The impact on the classification results will be discussed in the limitations section.
\renewcommand{\arraystretch}{1.0}
\begin{table}[htpb]
\begin{tabularx}{\linewidth}{XXXXXX}
\toprule
\textbf{Model}                              & \textbf{Condition}  & \textbf{Precision} & \textbf{Recall}  & \textbf{F-Measure} & \textbf{95\%-CI}               \\ \toprule
\multirow{3}{*}{GaussianNB}        & no context & 19.20\%   & 60.72\% & 23.68\%   & {[}20.44\%;26.91\%{]} \\
                                   & 1 prev turn  & 19.14\% & 60.67\% & 25.21\% &{[}21.96\%;28.47\%{]} \\
                                   & all prev turns & 14.09\% & 45.99\% & 19.51\%  & {[}15.61\%;23.41\%{]} \\ \midrule
\multirow{3}{*}{Random Forest}     & no context & 34.97\% & 13.02\% & 18.30\%   & {[}14.44\%;22.17\%{]} \\
                                   & 1 prev turn & 27.51\% & 04.35\% & 07.21\% & {[}05.12\%;09.30\%{]}\\
                                   & all prev turns & 02.79\% & 03.86\%  & 02.69\% & {[}01.51\%;03.87\%{]}   \\ \midrule
\multirow{3}{*}{SVM}               & no context & 32.41\% & 14.51\% & 19.13\% & {[}13.92\%;24.34\%{]}  \\
                                   & 1 prev turn  & 25.50\% & 07.11\%  & 10.49\% &{[}07.34\%;13.63\%{]}   \\
                                   & all prev turns & 00.16\% & 01.18\%  & 00.11\%  & {[}00.00\%;00.38\%{]}                    \\ \midrule \midrule
\multirow{3}{*}{mBERT}            & no context & 34.42\% & 36.84\% & 34.18\% & {[}27.51\%;40.84\%{]} \\
                                   & 1 prev turn & 33.76\% & 38.56\% & 34.46\% & {[}27.85\%;41.06\%{]} \\
                                   & all prev turns  & 21.26\% & 32.09\% & 22.82\% &{[}17.59\%;28.05\%{]} \\ \midrule
\multirow{3}{*}{XLM-RoBERTa}       & no context & 26.58\%   & 34.92\% & 27.61\%   & {[}21.01\%;34.20\%{]} \\
                                   & 1 prev turn  & 19.32\%   & 29.94\% & 20.36\%   & {[}14.70\%;26.01\%{]} \\
                                   & all prev turns & 15.37\% & 33.59\% & 18.64\%  & {[}14.18\%;23.09\%{]}  \\ \midrule
\multirow{3}{*}{GermanBERT}        & no context & 41.95\%   & 43.14\% & 41.43\%   & {[}34.57\%;48.28\%{]} \\
                                   & 1 prev turn  & 42.28\%   & 43.71\% & 42.11\%   & {[}35.19\%;49.03\%{]} \\
                                   & all prev turns  & 38.05\% & 40.88\% & 37.87\% & {[}30.71\%;45.03\%{]} \\ \bottomrule
\end{tabularx}
    \caption{Experiment results after 10-fold cross validation grouped by models and conditions.}
    \label{tab:exp_results}
\end{table}

\renewcommand{\arraystretch}{1.0}
\begin{table}[htpb]
\begin{tabularx}{\linewidth}{XXX}
\toprule
\textbf{Model} & \textbf{F-Measure} & \textbf{95\%-CI} \\
\toprule
GaussianNB & 22.80\% & {[}20.80\%;24.80\%{]} \\
\midrule
mBERT & \textbf{30.48\% ***} & {[}26.89\%;34.08\%{]} \\
XLM-RoBERTa & 22.20\% & {[}18.95\%;25.45\%{]} \\
GermanBERT & \textbf{40.47\% ***} & {[}36.48\%;44.46\%{]} \\
\bottomrule
\end{tabularx}
\caption{Overall average F1 scores model performance. In bold the significant results against GNB, with *** : $p<0.001$ }
\label{tab:baseline_bert_comparison}
\end{table}

\subsubsection{Results}
The results are illustrated in Table \ref{tab:exp_results}. We report precision, recall and the F-measure. Accuracy was not taken into consideration since it would not be a reliable measure for such imbalanced datasets. In the following paragraphs, we assess the performance of the different models that were employed. In the subsequent sections, we present the results that were achieved regarding the different context conditions and information needs.
\paragraph{Model performance:}
In Table \ref{tab:exp_results}, we report the values achieved when removing stopwords for classification. While no significant difference could be observed when stopwords were removed for the Random Forest and SVM classifiers, removing stopwords when using the GaussianNB classifier significantly improved classification performance ($p < 0.05$) compared to the results where stopwords are kept. GaussianNB also outperforms Random Forest and SVM regarding the F-Measure in detecting information needs ($p<0.001$ for both) and, thus, will be the baseline to test against BERT-based models.

To assess the overall model performance, a one-way ANOVA as well as post-hoc tests with bonferroni-adjusted values were conducted. These indicate that the GermanBERT model performs significantly better than mBERT ($p<0.001$) and XLM-RoBERTa ($p<0.001$). Compared to the baseline model, mBERT and GermanBERT perform significantly better ($p<0.001$). However, no difference was found between XLM-RoBERTa and GaussianNB.

Fine-tuning BERT with a language specific model, i.e. GermanBERT, on the text classification task yielded strong results compared to previous approaches with word embeddings that were, for example, employed in Frummet et al. \cite{Frummet2019}. The main reason for this is that BERT can much better capture contextual information than for instance word2vec or FastText embeddings can \cite{devlin2019bert}. Among the BERT models we tested, GermanBERT was found to yield the best results. One reasonable explanation for this is the more accurate tokenization by GermanBERT which leads to better understanding of words by the BERT model\footnote{see https://www.deepset.ai/german-bert}.

\paragraph{Context Conditions:}
\begin{figure}
    \includegraphics[width=\linewidth]{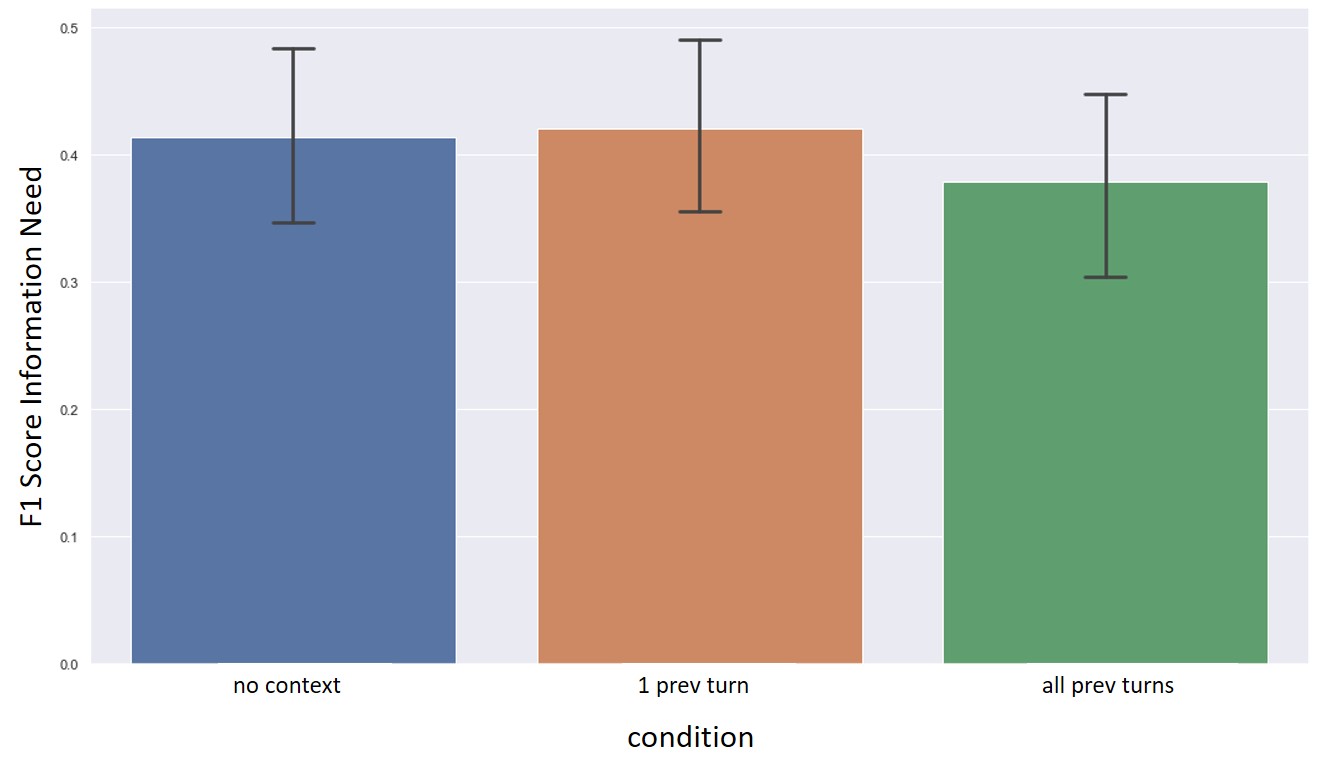}
    \caption{Average F1 Score for each condition over all information needs using the GermanBERT model.}
    \label{fig:context_condition_results}
\end{figure}
\begin{figure}
    \includegraphics[width=\linewidth]{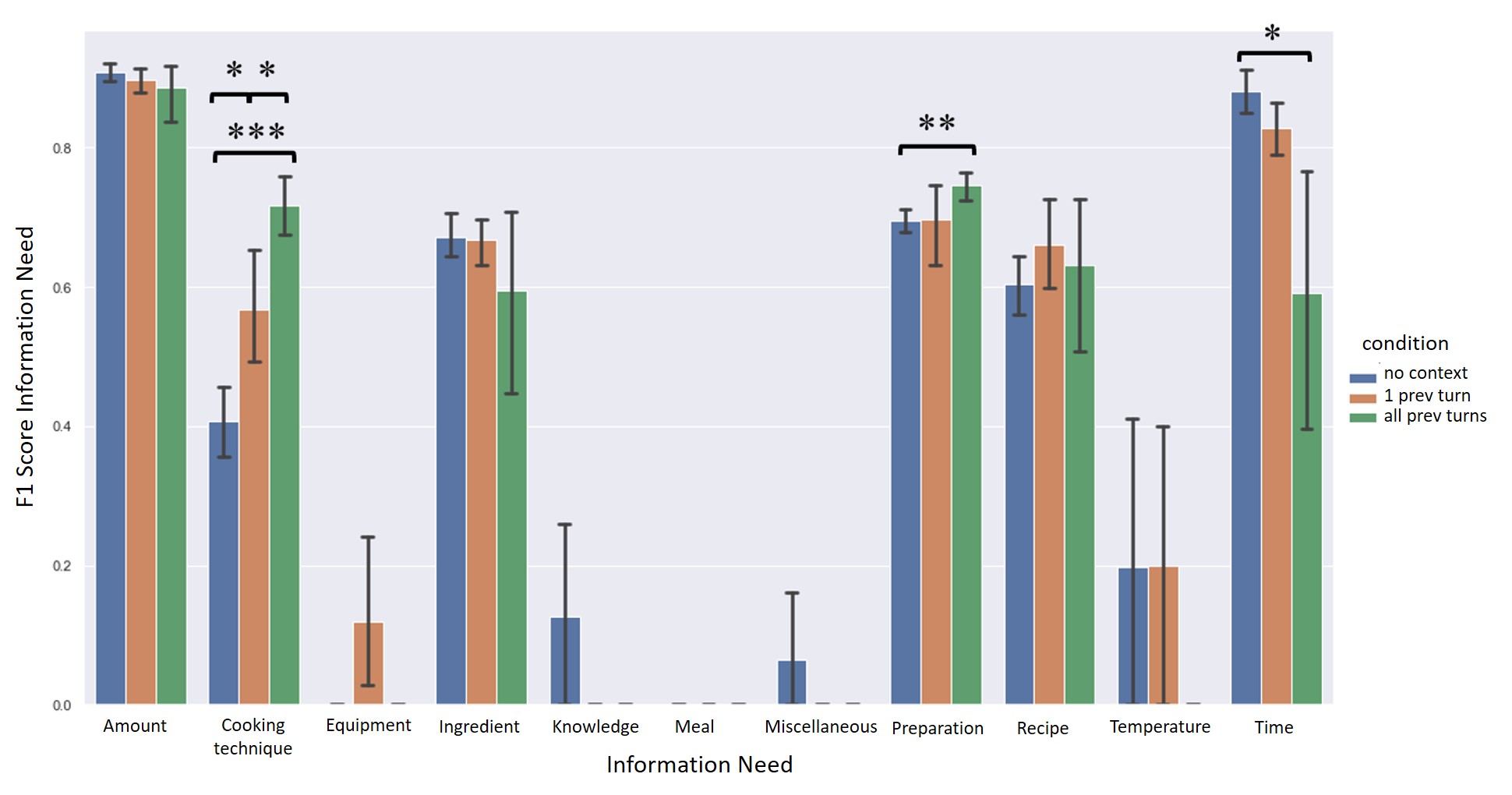}
    \caption{F1 Score for each information need grouped by condition using the GermanBERT model, with * : $p<0.05$;  ** : $p<0.01$;  *** : $p<0.001$.}
    \label{fig:information_needs_results}
\end{figure}

Since the GermanBERT model performs best, this model was employed to further assess the classification performance regarding context conditions and information need prediction. For this, Figure \ref{fig:context_condition_results} illustrates that while the \textit{1 prev turn} condition  ($M=42.11\%$) performs slightly better than \textit{no context} ($M=41.43\%$) and \textit{all prev turns} ($M=37.87\%$), no significant differences were detected after conducting one-way ANOVA analysis ($F=0.42, p=0.69$). The average F1 score achieved for GermanBERT among all conditions was \textbf{$40.47\%$}.

\paragraph{Information Need prediction:}
The achieved F1 scores, grouped by information need and condition for the GermanBERT model, as well as an overview of which conditions lead to significantly better results within information need categories are illustrated in Figure \ref{fig:information_needs_results}.

While some information need categories, e.g., \textit{Amount} and \textit{Time}, performed quite well, \textit{Equipment}, \textit{Knowledge}, \textit{Meal}, \textit{Miscellaneous} and \textit{Temperature} achieved F1 scores below 20\%. This is likely due to the lack of data for these categories since the aforementioned needs only relate to, on average, 35 turns. While BERT can usually deal quite well with limited data \cite{madabushi2019costbert}, limits seem to be reached with this dataset. The lack of data for these categories mean that the results achieved cannot be interpreted reliably. We will, therefore, focus on the interpretation of the remaining information need categories where it is possible to identify several patterns.

For the fact-based information needs \textit{Amount}, \textit{Time} and \textit{Ingredient}, the no context condition performs best. The results of the \textit{Time} information need indicate a significantly better performance for the condition where no context information was taken into account compared to the \textit{all prev turns} condition. This aligns with impression gained during the qualitative analysis that annotating these needs was straightforward as little context needs to be accounted for. 
Similar patterns were observed for the competence-oriented information needs \textit{Cooking technique} and \textit{Preparation}. For both needs, the \textit{all prev turns} condition performed significantly better than the no context condition. Turns labelled as \textit{Cooking technique} performed significantly better when more context was modelled. The \textit{1 prev turn} condition showed significant better results compared to the \textit{no context} one and \textit{all prev turns} showed significant better performance than \textit{1 prev turn}. This, again, aligns with the findings from Section \ref{sec:in_communication}.

The classification at level 1 gives the impression that fact-based needs are easier to predict, which aligns with the experiences of the manual annotation as explained in Section \ref{sec:in_description}. To test this, we formulated a similar prediction task with the same setup, but for level 0 needs. We employed the GermanBERT model and fine-tuned this as it was done in the best performing model in the first round of experiments.
Fact-based information needs are indeed significantly easier to detect than competence needs ($M_{fact}=86.68\%$, $M_{competence}=74.78\%$, $t=4.20, p<0.001$).
~\\
~\\
Our experiments demonstrate that fact-based needs such as \textit{Amount} and \textit{Time} can be predicted with high confidence and require little context information. On the other hand, classifiers for competence oriented needs, i.e. \textit{Cooking technique} and \textit{Preparation}, perform better when more context information is added, but are generally less accurate than classifiers for fact-based needs.

\section{Discussion}
\label{sec:discussion}
In this section, the results reported above will be discussed. First, in Section \ref{sec:comparison_cooksearch_other}, we focus on our information need taxonomy. We compare it to other coding schemes from the literature and illustrate the additional value that is offered by our \textit{CookversationalSearch} taxonomy. Second, in Section \ref{sec:implication_conv_ass}, we discuss the implications our results have on the design of conversational assistants and extrapolate what needs to be considered when implementing a conversational search system. Finally, we elaborate on the relevance of context for information need classification and discuss why some needs are easier to predict than others.
\subsection{Comparing CookversationalSearch with other conversational taxonomies}
\label{sec:comparison_cooksearch_other}
To assess the value of our dataset and our taxonomy for the research community, we discuss differences to cooking and non-cooking related taxonomies and annotation schemes.
\subsubsection{Comparison with domain-agnostic taxonomies}
A prominent domain-agnostic taxonomy is the Dialogue Act ISO annotation scheme that was introduced by Bunt et al. \cite{Bunt2017dialgueact}. This hierarchical annotation scheme contains, at the highest level, 26 communicative functions that can be used to label turns, e.g., \textit{Accept Suggest} and \textit{Decline Suggest}. Some of these labels relate to codes in our taxonomy. For example, the \textit{Accept Suggest} move can be observed in the \textit{Recipe Selection} need since participants can accept or reject suggested recipes. Other similar dialogue related functions can also be observed in datasets, such as VERBMOBIL \cite{jekat1995verbmobil} where codes, such as  \textit{Clarify} are assigned to turns that aim to make taken actions explicit and seek reassurance. \textit{Reassurance}s form part of our information need taxonomy and this relates to the \textit{Clarify} dialogue act, which is part of Bunt et al.'s taxonomy, too. A difference between these domain-agnostic coding schemes and the one we present is that these lack a description of the associated content. Knowing what is being reassured would help the system to provide better assistance in the sense that systems then know how they should react. For example, \textit{Amount of Ingredient - Reassurance} (level 3 code) specifies that participants want to be reassured with respect to the amount of a certain ingredient. A system can then react accordingly and clarify if the amount of an ingredient has been understood correctly.

Our coding scheme also goes beyond that proposed by Shiga et al. \cite{Shiga2017} which, in contrast to the aforementioned coding schemes, offers a stronger relation to  information needs. They employed Taylor's information needs \cite{Taylor1968} and combined them with high-level task related needs, such as \textit{Topic}, \textit{Problem Solving} and \textit{Search} in a travel planning task. Although - like our work - the coding scheme is applied in a conversational search setting, the way the needs are classified was coarse-grained and abstract. We argue that these labels are not specific enough to adequately describe travel planning content such that systems can provide assistance. Our taxonomy, on the other hand, illustrates that two different kinds of assistance should be provided, namely fact-based, where information is provided, and competence-oriented, which revolves around teaching skills. Moreover, the tree-like structure allows needs to be distinguished based on content. This, again, helps the system tailor the information provided.

To summarise, the coding scheme we developed offers utility beyond domain-agnostic taxonomies (see e.g. Bunt et al. \cite{Bunt2017dialgueact}) and the information need classifications (e.g. Taylor \cite{Taylor1968} and Shiga et al. \cite{Shiga2017}) since it offers finer grained descriptive qualities that are content specific. In the next chapter, we relate to our contribution to previously published taxonomies related to cooking.

\subsubsection{Comparison with cooking-related taxonomies}
Our coding scheme goes beyond those previously collated in the cooking domain. For example, Nouri et al. \cite{nouri2019supporting} created a very controlled environment and had a very small group of users (eight participants) that used smart speakers privately and were cooking pre-selected recipes. In a similar environment, Vtyurina \& Fourney \cite{vtyurina2018} analysed intents that occur during cooking, again with only 10 participants. In our study, we had 45 participants and the recipe search was part of the natural dialogue (i.e. uncontrolled). Participants in our study moreover conversed in a more natural manner with the system, i.e. the experimenter. Due to the small sample size, Nouri et al. only identified a small number of cooking-related intents, which we list in Table \ref{tab:info_need_comparison_nouri_cunningham}. Table 7 shows how the level 1 categories in our taxonomy relate to those reprorted in Nouri et al. and Cunningahm \& Bainbridge. For intents like ingredient\_quantity, ingredient\_list\_all and step\_next, \textit{Amount}, \textit{Ingredient} and \textit{Preparation} can be found as equivalents. Our taxonomy, however, goes beyond their limited list of cooking-related intents to offer a more broader and finer-grained description of needs that can occur during cooking.
\renewcommand{\arraystretch}{1.0}
\begin{table}[htpb]
\small
\begin{tabularx}{\linewidth}{sXs}
\toprule
\textbf{Intents \scriptsize{(Nouri et al.)}} & \textbf{Response Types \scriptsize{(Cunningham \& Bainbridge)}} & \textbf{Level 1 info. need}  \\
\toprule
ingredient\_quantity & -- & Amount\\
ingredient\_list\_all & -- &  Ingredient\\
step\_next & -- & Preparation\\
recipe\_name & Recipes &  Recipe\\
 -- & Definitions & Knowledge  \\
 -- & Cooking Techniques & Cooking technique \\
 -- & -- & Time \\
 -- & -- & Equipment \\
\bottomrule
\end{tabularx}
\caption{Comparison of selected cooking user intents and needs from Nouri et al. \cite{nouri2019supporting}, Cunningham \& Bainbridge \cite{CunninghamBainbridge2013} and our work (middle column). For demonstration purposes, we highlight just some of the categories that are in our taxonomy and not in past works.}
\label{tab:info_need_comparison_nouri_cunningham}
\end{table}

A second obvious point of comparison is the taxonomy created by Cunningham \& Bainbridge \cite{CunninghamBainbridge2013}, which is perhaps the closest to our work. As outlined in Section \ref{sec:related-work}, their taxonomy is based on Google Answer posts that were created by hobby chefs and distinguishes 13 different needs that were uttered in these posts. Most frequently, posters requested information about products and recipes, but were also curious about definitions and cooking techniques. These are needs similar to the ones detected in our corpus, i.e. \textit{Recipe}, \textit{Knowledge} (akin to \textit{definitions}) and \textit{Cooking technique}. However, Cunningham \& Bainbridge's taxonomy covers only needs of one specific group, i.e. hobby chefs, which is the reason why needs like food trivia and food history, which are less likely to occur in an everyday cooking situation, are part of their taxonomy. Our coding scheme is more applicable to practical situations where assistance may be required, which is reflected in additional codes such as \textit{time} being prominent.


This section has illustrated the value of our taxonomy for the research community. It is the first conversational search related taxonomy for the cooking domain that represents information needs to such a level of detail and, at the same time, contains codes that allow a broader range of needs - ones that are likely to require assistance in everyday cooking - to be described.

\subsection{Implications on the interaction design of conversational assistants}
\label{sec:implication_conv_ass}
In this section, we discuss which aspects are considered relevant when designing the interaction with conversational assistants. With this aim in mind, we  elaborate on the role of conversational cues in conversational search dialogues, as well as to which extent our findings may be generalised beyond the scenario of assisting cooking.

\subsubsection{The role of conversational cues}
\label{sec:role_conversational_cues}
In Section \ref{sec:in_communication}, we identified conversational cues, which seem to be relevant in the context of information need communication and, thus, should be considered when designing conversational systems. We observed the use of both explicit and implicit queries to convey information needs. Conversational agents often only process turns that exhibit the shape of explicit queries (see \cite{Kopp2005, Allen2001}) as is the case in the example: \enquote{Which ingredients are needed?}. Queries of this type make it easier for systems to interpret utterances as system input that needs to be processed since less context modeling is required \cite{ren2020}. Existing dialogue management systems, such as \textit{RavenClaw} \cite{bohus2009ravenclaw}, which are employed in task-oriented dialogue contexts, are good at handling explicit requests, they have trouble coping with the implicit queries. If such systems are to interact as humans do, however, they would need to deal with the kind of implicit requests that were outlined in Section \ref{sec:in_communication}. Our results align with findings from Vtyurina \& Fourney with both studies finding a similar proportion ($\approx 25\%$) of implicit queries.
In their analyses, Vtyurina \& Fourney \cite{vtyurina2018} also found that grounding behaviour was exhibited by means of verbatim repetitions of what has been understood -- a pattern described as \textit{Reassurance} in our study. For example, with a turn like \enquote{Covered?} (part. 15), a participant is seeking reassurance if the taken action is correct and, therefore, grounds their behaviour by requesting feedback from the system. Grounding and relevance feedback have also been considered appropriate for conversational systems by Trippas et al. \cite{Trippas2020}.


~\\
To summarise, our findings reinforce patterns observed in previous studies in that they underline - even in this specific cooking context context - that conversational cues are important for assistants to identify and provide appropriate assistance. Accounting for concepts such as implicit queries, grounding and reassurance are imperative if human-like interaction patterns are to be supported.

\subsubsection{Generalisability of our findings}
The presented study was performed in the fairly specific context of home cooking. Nevertheless, there are aspects of the task performed that might mean that our findings transfer to other situations. Cooking tasks are highly procedural, goal-oriented tasks. They involve completing  a sequence of individual steps to achieve a larger goal. There are other situations which would be comparable. For example, we could imagine tasks such as bicycle maintenance, e.g. fixing a punctured tyre, or building flat-pack furniture would have considerable overlap with the cooking examples studied here. We believe, for instance, that since such tasks are similarly sequential in one could expect similar usage of discourse markers like \enquote{Mhm.} or \enquote{Ok.} indicate task completion.

We would, furthermore, anticipate a similar high-level delineation of information need types in these situations to fact-based / competence-based distinction we present at level-0 in our taxonomy. For example, when building a shelf, a user might ask \enquote{How many screws do I need to fix the bottom plate?}(fact-based). On the other hand, competence-oriented needs could be also be expected, e.g. \enquote{How do I fix the bottom plate to the wall?}(competence-oriented). 

We would argue that it is not only the high-level codes that would transfer. It seems plausible that some level 1 needs may also apply in other domains where procedural tasks take place. The turn \enquote{How many screws do I need to fix the bottom plate?}, for example, expresses an Amount related need that can be used in both the cooking domain and the DIY domain for instance. Also, utterances asking for system capabilities, such as \enquote{Do you have an integrated timer?} (labelled as Miscellaneous), seem to be generic and applicable to other domains. Similarly, turns like \enquote{What do I have to do next?}, which are labelled as Preparation in our corpus, seem to generalise beyond the cooking-context.

One could easily perform a comparable study in a DIY context to test to what extent our codes and also the classification performance transfer to this domain. The Alexa Prize TaskBot challenge 2021\footnote{see https://www.amazon.science/academic-engagements/ten-university-teams-selected-to-participate-in-alexa-prize-taskbot-challenge}, where teams develop conversational assistants for both cooking and DIY tasks, underlines how closely related to each other these domains are.

In this section, we have discussed the relevance of conversational cues for system design and arguing, for example, that implicit requests are important, particularly in procedural tasks, to replicate human to human style interaction. We have, moreover, argued that there is potential for our results to transfer into other domains that feature procedural tasks. We named DIY assistance is one example.
\subsection{On the relevance of context}
\label{sec:context_relevance}

Both the qualitative annotation process and the prediction experiments underlined the importance of context information needs. The results from our prediction experiment revealed that fact-based needs perform significantly better than competence oriented needs. One reason for this is that fact-based information needs contain more information need discriminating words (illustrated in Table \ref{tab:top15_words}). The annotation process revealed that the competence oriented categories require more context to be taken into account given the lack of distinctive features, i.e. words, which makes it challenging for both human coders and classifiers to distinguish classes. This assertion is supported by the classification experiment results. While fact-based needs, such as \textit{Amount} and \textit{Time}, achieved F1 scores  $\approx90\%$, \textit{Cooking technique} and \textit{Preparation} achieved results $\approx70\%$ in their best performing condition. Including the context in the form of previous turns, however, significantly improved results in competence-oriented needs \textit{Cooking technique} and \textit{Preparation}. 
~\\
With respect to the development of future systems, we argue that while we can reliably detect Amount and Time needs and, to some extent, other fact-based needs, such as Temperature, which achieved low results due to lack of data, other strategies need to be established to identify competence needs. Similar to our results, Ren et al. \cite{ren2020} and Aliannejadi et al. \cite{aliannejadi2019} showed that including more context, i.e. conversational history, improves the classification performance of the current turn. Recently, this problem has been investigated by Nouri et al. \cite{nouri2020} who tried to enable step-wise recommendations for current steps in a controlled cooking task including different contextual information.
~\\
Our results demonstrate the feasibility of predicting some types of need. To further improve the detection of the trickier to identify classes, we will need to find other ways for augmenting and representing the cooking process and the context. One way to achieve this may be to reformulate queries by adding missing context from the conversation history to the current turn query \cite{voskarides2020queryresolution}. In our study, we employed a simple approach, modelling information need turns with different amounts of context by adding previous queries as plain text to the current query. However, more elaborate methods exist to better model context, see, for example, Voskarides et al. \cite{voskarides2020queryresolution}, Yu et al. \cite{yu2020fewshot} and Lin et al. \cite{Lin2020QueryRU,Lin2020ConversationalQR}. In future work, more effort can be put into this by building on prior work.
~\\
~\\
In summary, our results -- both of the annotation and classification highlight the importance of context, particularly in identifying competence needs. Applying more elaborate strategies to achieve a better representation of context information will be part of our future work.
\section{Limitations}
\label{sec:limitations}
In this section, we reflect on some of the limitations of this study.
When performing data collection with 45 participants we cannot exclude learning effects by the experimenter mimicking a conversational system. From session to session, the experimenter increased in confidence and developed strategies for interacting with participants. This affected, for example, the amount of information that was provided when participants issued queries. The means of interacting with participants was certainly not completely consistent throughout the study. We do not believe, however, that this detracts from the quality of the data since the interaction patterns were always plausible, human dialogue. Moreover, we did not observe different information needs or particular phraseology at different time points when transcribing or analysing the data. It is also possible that having the experimenter act as the assistant led to specific outcomes (Experimenter bias). However, this is unlikely as the experimenter did not have any pre-conceived ideas of what assistance should be provided and the conversations were largely driven by the participants with utterances arising naturally from the cooking process.
~\\
A further point worthy of discussion is that since we conducted an in-situ study, both experimenter and participant were able to see and interact with each other using non-oral communication (e.g. body language). This raises the question if the physical environment, i.e. gesutres and visual apsects, might have influenced the dialogue interaction. While we fully relied on speech-only interaction and the experimenter did not accept queries that were not possible in such a scenario, minor environmental effects cannot be fully excluded. When participants employed anaphoras to refer to e.g. ingredients like in \enquote{Is \textbf{this} enough?} seeing what the participant was doing might have, unconsciously, simplified the interpretation of utterances which would be less clear in spoken-only wizard-of-oz scenario.
~\\
One limitation of our quantitative experiment relates to the representation of the multi-label classification as binary classification task. While we learned about how well information needs can be identified based on an utterance, we cannot measure whether co-occurence of certain needs influence one another, which is a common issue for binary relevance classification (see Luaces et al. \cite{Luaces2012BinaryRE} and Zhang et al. \cite{Zhang2017BinaryRF}). The influence of co-occurring information needs on the classification results can be investigated in future studies.
~\\
Finally, when modelling context, we employed the plain text from previous turns. This text can, however, be represented in more elaborate ways as outlined in Section \ref{sec:implication_conv_ass}. Moreover, the impact of linguistic features, such as intonation and the explicit-implicit dimension on the classification results, could have been investigated in further detail and will be part of our future work.

\section{Conclusion}
\label{sec:conclusion}
Our results provide an in-depth understanding of the information needs occurring in the domain of home cooking. RQ1 rasied the question of what kind of information needs arise when cooking with a conversational assistant. To answer this we provided a detailed hierarchical information need taxonomy that shows us a variety of different needs that can occur in a cooking scenario. These needs can broadly be categorised in different types, i.e. fact-based and competence-oriented, which require different types and degrees of assistance. Addressing RQ2, these needs can be communicated through different linguistic means, both as explicit and implicit queries, and discourse markers are often used. Furthermore, our findings indicate the feasibility of identifying needs automatically (RQ3). In this respect, our experiments demonstrate some success with fact-based information needs being particularly feasible. We have also shed light on the problems that need to be solved to successfully detect and process other, more challenging information need types in an automated scenario. These mainly include finding methods to better incorporate context and conversational history. Future work will focus on creating methods to better incorporate context information in the classification and assistance process.

\bibliographystyle{abbrv}
\bibliography{paper_arxiv}  






\end{document}